\documentclass[aps,twocolumn]{revtex4}
\makeatletter

\newcommand{\Rmnum}[1]{\expandafter\@slowromancap\romannumeral #1@}
\makeatother
\usepackage{graphicx}
\usepackage{dcolumn}
\usepackage{bm}
\usepackage{CJK}
\usepackage{color}
\usepackage{amsfonts}
\usepackage{psfrag}
\usepackage{wrapfig}
\usepackage{subfigure}
\usepackage{makeidx}
\usepackage{multirow}
\usepackage{epsf}
\usepackage[colorlinks,linkcolor=red,anchorcolor=red,citecolor=blue,urlcolor=blue]{hyperref}
\usepackage{amsmath}
\usepackage{cases}
\usepackage{bm}
\usepackage[dvipsnames]{xcolor}

\begin{document}
\begin{CJK}{GBK}{song}
\title{Higher-order modulation instability and multi-Akhmediev breathers of Manakov equations: Frequency jumps over the stable gaps between the instability bands}
\author{Shao-Chun Chen$^{1}$}
\author{Chong Liu$^{1,2,3,4}$}\email{chongliu@nwu.edu.cn}
\author{Nail Akhmediev$^{2,5}$}%\email{Nail.Akhmediev@anu.edu.au}
\address{$^1$School of Physics, Northwest University, Xi'an 710127, China}
\address{$^2$Department of Fundamental and Theoretical  Physics, Research School of Physics, The Australian National University, Canberra, ACT 2600, Australia}
\address{$^3$Shaanxi Key Laboratory for Theoretical Physics Frontiers, Xi'an 710127, China}
\address{$^4$NSFC-SPTP Peng Huanwu Center for Fundamental Theory, Xi'an 710127, China}
\address{$^5$Arts $\&$ Sciences Division, Texas A$\&$M University at Qatar, Doha, Qatar}

%%%%%%%%%%%%%%%%%%%%%%%%%%%%%%%%%%%%%%%%%%%%
\begin{abstract}
We study higher-order modulation instability phenomena in the frame of Manakov equations.
Evolution that starts with a single pair of sidebands expands over several higher harmonics. The choice of initial pair of sidebands influences the structure of unstable frequency components and changes drastically the wave evolution leading, in some cases, to jumps across spectral components within the discrete spectrum.
This complex dynamics includes several growth-decay cycles of evolution.
We show this using numerical simulations of the MI process and confirm the results using the exact multi-Akhmediev breather solutions.
Detailed explanation of the observed phenomena are given.
\end{abstract}

\maketitle
\section{Introduction}

Modulation instability (MI) is a phenomenon well known in optics \cite{Bespalov}, hydrodynamics \cite{Benjamin} and and other branches of physics. Simply speaking, MI is an instability of a plane wave or a continuous wave in a self-focusing nonlinear medium relative to periodic modulations. Small modulations with frequencies  within the band of instability are amplified. However, exponential amplification is valid only at the initial stage of the process that is usually described by the linear stability analysis. Once the initial stage is developed, further evolution is becoming more complex. There are several reasons for complications. Firstly, with the nonlinear growth of the amplitude of modulation the spectrum of the wave field expands due to the four-wave mixing process. The full spectrum of the wave field may increase far beyond the instability band \cite{Exp2009}. Secondly, the higher harmonics of modulation have also a chance to be amplified provided that they are located within the instability band \cite{AEK}.
These reasons lead to a more complex evolution of the wave field.

In the simplest case when only one pair of sidebands is involved in the dynamics results in the   excitation of `Akhmediev breathers' (AB) \cite{AB}. This is a special solution of the focusing nonlinear Schr\"odinger equation (NLSE) that describes the full growth-decay cycle of the periodic perturbation on top of a plane wave \cite{AB,TMP1987,MC1}. The case, when several unstable spectral components are involved in the evolution is known as `higher-order' modulation instability \cite{OC2010,Exp2011-PRL,Exp2017-PRE}.
In these situations, several ABs can be excited simultaneously.
Then the full evolution can be described by the multi-AB solutions of the NLSE \cite{JETP88}. Each AB that is involved in the dynamics expands a pair of sidebands of initial modulation to a whole set of frequency components that are the higher harmonics of initial sidebands. The resulting wave field  depends on how many of these frequency components fall within the instability spectrum. In the case of systems described by the NLSE, even such complicated dynamics can be described analytically although comparison with the results of an experiment or numerical simulations can be quite involved and requires a deeper analysis.

In the case of the NLSE, there is a single band of unstable frequencies thus making such analysis relatively easy task. In more complicated cases, several instability bands may exist \cite{VMI1987,VMI1991,VMI2014,DF4}. In particular, this happens when several wave components are involved in the dynamics \cite{OF,BEC,F}.  The MI of the coupled evolution equations has been studied in several publications \cite{VMI2013,VMI2015,VRW2016,VRW2018,VRW2017,VRW2022}. However, the role of additional MI bands in the nonlinear stage of MI evolution remains largely unexplored.

Interaction between multiple waves is one of the reasons for appearance of new MI bands. For example, in the case of Manakov equations \cite{Manakov1974}, there is an additional `X-shaped' MI band well known from the previous studies \cite{VMI2000}. Thus, when the spectrum expanding as a result of MI covers these additional bands, the wave evolution may enter new regimes unknown for the case of the NLSE. Moreover, the presence of spectral gaps between the bands of unstable frequencies may introduce additional complications into the dynamics.
Recent results revealed the existence and the extremely asymmetric spectra of fundamental ABs growing in such MI bands \cite{VAB2021,VAB2022}. The spectral expansion in these cases may traverse the gaps thus causing jumps from lower spectral harmonics to higher-order ones  skipping some intermediate frequencies. This is one of the new complex phenomena related to the MI that we observed in the present study.

In this paper, we study numerically wave evolution caused by the MI with a single pair of sidebands and observe its evolution when the initial parameters of modulation vary.
We did indeed observed the regime when the spectral expansion occurs through the jumps over the spectral components that fall within the stability gap. Detailed explanation of this phenomenon is presented and exact multi-AB solutions are given that correspond to such phenomena.

% Section II
\section{Vector ABs and MI}\label{Sec2}

The Manakov equations \cite{Manakov1974}, in dimensionless form, are given by
\begin{equation}\label{eq1}
\begin{split}
i\frac{\partial\psi^{(1)}}{\partial t}+\frac{1}{2}\frac{\partial^2\psi^{(1)}}{\partial x^2}+\sigma(|\psi^{(1)}|^2+|\psi^{(2)}|^2)\psi^{(1)}&=0,\\
i\frac{\partial\psi^{(2)}}{\partial t}+\frac{1}{2}\frac{\partial^2\psi^{(2)}}{\partial x^2}+\sigma(|\psi^{(1)}|^2+|\psi^{(2)}|^2)\psi^{(2)}&=0,
\end{split}
\end{equation}
where $\psi^{(1)}(t,x)$, $\psi^{(2)}(t,x)$ are the two nonlinearly coupled components of the vector wave field. The physical meaning of independent variables $x$ and $t$ depends on a particular physical problem of interest.
We have normalised Eq. (\ref{eq1}) in a way such that $\sigma=\pm1$.
Note that in the case $\sigma=1$, Eqs. (\ref{eq1}) describe the focusing
(or anomalous dispersion) regime, In the case $\sigma=-1$,
Eqs. (\ref{eq1}) describe the defocusing (or normal dispersion) regime.

%\subsection{Vector AB solutions}\label{Sec2.1}
We start with the fundamental AB solution of (\ref{eq1}).
Using a Darboux transformation scheme for Eqs. (\ref{eq1}) and using the vector plane wave solution of Eqs. (\ref{eq1})
\begin{equation}
\psi_{0}^{(j)}=a \exp \left\{ i \left[{\beta_j}x + (2 \sigma a^2- \beta_j^2/2) t \right] \right\},\label{eqpw}
\end{equation}
with $j=1, 2$ as a seed, at the first step, we find \cite{VAB2021,VAB2022}:
\begin{equation}
\psi^{(j)}=\psi_{0}^{(j)}\left[\frac{\cosh(\bm{\Gamma}+i\gamma_j)e^{i\eta_{1j}}+\varpi \cos(\bm{\Omega}-i\epsilon_j)e^{i\eta_{2j}}}{\cosh\bm{\Gamma}+\varpi \cos\bm{\Omega}}\right].\label{eqb}
\end{equation}
Parameters $a$ and $\beta_j$ in (\ref{eqpw}) are the amplitude and the wavenumber of the two plane wave components respectively. Without loss of generality, we can set $\beta_1=-\beta_2=\beta$.
The scalar arguments $\bm{\Gamma}$ and $\bm{\Omega}$ in (\ref{eqb}) are:
\begin{eqnarray}
\bm{\Gamma}=\omega \bm{\chi}_i\bm t,~
\bm{\Omega}=\omega \left[\bm x+ (\bm{\chi}_r+\frac{1}{2}\omega )\bm t \right]+\arg\frac{2\bm{\chi}_i}{2\bm{\chi}_i-i\omega}.
\end{eqnarray}
Here $\bm{x}=x-x_{1}$, $\bm{t}=t-t_{1}$ are shifted spatial and time variables respectively with  $x_{1}$ and $t_{1}$ being responsible for the spatial and temporal position of the centre of the breather.
Other notations in (\ref{eqb}) are:
\begin{eqnarray}
\eta_{1j}&=&\frac{\gamma_{1j}+\gamma_{2j}}{2},~~\eta_{2j}=\arg\frac{\bm{\chi}^*+\beta_j}{\bm{\chi}+\beta_j+\omega}, \\
\gamma_{j}&=&\frac{\gamma_{1j}-\gamma_{2j}}{2},~~\varpi= \Big|\frac{2\bm{\chi}_i}{2\bm{\chi}_i+i\omega} \Big|,\\
\gamma_{1j}&=&\arg\frac{\bm{\chi}^*+\beta_j}{\bm{\bm\chi}+\beta_j},~~
\gamma_{2j}=\arg\frac{\bm{\chi}^*+\beta_j+\omega}{\bm{\bm\chi}+\beta_j+\omega},\\
\epsilon_j&=&\log\left(\frac{(\bm{\chi}^*+\beta_j)(\bm{\chi}+\beta_j)}{(\bm{\chi}+\beta_j+\omega)(\bm{\chi}^*+\beta_j+\omega)}\right)^{1/2}.
\end{eqnarray}

% Figure 1
\begin{figure}[htb]
\centering
\includegraphics[width=86mm]{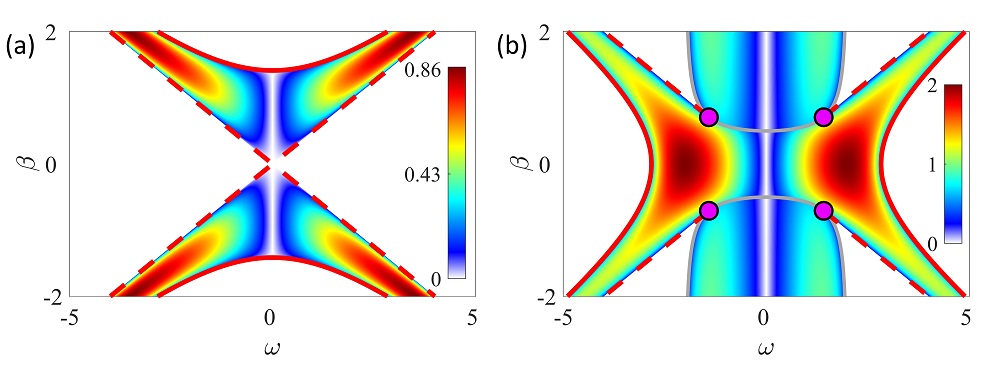}
\caption{MI growth rate $G=|{\omega} \bm{\bm\chi}_i|$ on the ($\omega,\beta$) plane given by the AB solution (\ref{eqb}) in (a) defocusing and (b) focusing regimes. The four large violet dots in (b) are the branch points given by (\ref{eq-bp}).
Parameter $a=1$.}\label{f1}
\end{figure}

An important parameter of the breather is its complex eigenvalue $\bm{\chi}\equiv\bm{\chi}(\sigma, a, \beta, \omega)$ with its real $\bm{\chi}_r$ and imaginary $\bm{\bm\chi}_i$ parts. The explicit expressions for $\bm{\chi}$ are given by:
\begin{eqnarray}
\bm{\chi}_{\pm}=\pm(\beta^2-\sigma a^2+\omega^2/4-\sqrt{\bm{\nu}})^{1/2}-\omega/2,\label{eqchi}
\end{eqnarray}
where
\begin{eqnarray}
\bm{\nu}=a^4-4\sigma a^2\beta^2+\omega^2\beta^2.
\end{eqnarray}
Thus, once $\sigma$ is fixed, the AB (\ref{eqb}) is a three-parameter family of solutions depending on the background amplitudes $a$, the relative wavenumber $\beta$, and the modulation frequency $\omega$.
The solution (\ref{eqb}) represents the full growth-decay cycle of MI. Namely, it
grows out of the plane wave (\ref{eqpw}) that is weakly modulated  with frequency $\omega$.

The AB solution satisfies a simple transformation
\begin{eqnarray}
\psi^{(j)}(\bm x, \bm{\chi}_{+})=\psi^{(j)}(\bm x+\Delta x,\bm{\chi}_{-})e^{i\Delta \phi_j},\label{eqpin}
\end{eqnarray}
where $\Delta x=\frac{2}{\omega}\left(\arg\frac{2\bm{\chi}_i}{2\bm{\chi}_i-i\omega}\right)$ is the shift along the $x$-axis and $\Delta \phi_j=2\eta_{1j}$ is the phase shift of the complex function, respectively. This means that $\psi^{(j)}(\bm{\chi}_{+})$ and $\psi^{(j)}(\bm{\chi}_{-})$ have the same amplitude profiles.
This is essentially the same solution but shifted along the x-axis.
%This is an example of degeneracy when for any given initial parameters there are two solutions with the same profile for two different eigenvalues.

The growth rate of MI that follows from the exact AB solution is $G=|\omega \bm{\chi}_i|$.
The plane wave is unstable when $G\neq0$. Remarkably, the plane wave can be unstable both in the focusing and in the defocusing cases. The values of the MI growth rate on the ($\omega,\beta$) plane separately for the defocusing and focusing cases are shown in Figs. \ref{f1}(a) and \ref{f1}(b) respectively. The growth rate $G$ depends on both the frequency $\omega$ and the eigenvalue $\bm{\chi}$. The areas of nonzero growth rate in Figs. \ref{f1} are simultaneously the areas of the AB existence.

Let us analyse first the MI growth rate plot in the defocusing case, $\sigma=-1$, shown in Fig. \ref{f1}(a). The MI growth rate is positive in the X-shaped region located between the two red dotted straight lines $\omega^2=4\beta^2$ and the two red solid curves defined by $\omega^2=4(\beta^2-2a^2)$. $G$ is a continuous function of $\omega$ and $\beta$ within these areas.

The growth rate curve at the first sideband frequency defines further evolution of the perturbed wave field. In particular, the plots in Figs. \ref{f1} allow us to generalise the ideas of higher-order MI dynamics developed in Ref. \cite{DF4} to the vector wave field.
If $|\beta|\geq2 \sqrt6 a/3$, the second harmonic $2\omega$ is always outside the MI band. This means that higher-order MI cannot occur.
The higher-order MI can be excited with a single initial modulation frequency $\omega$, when $|\beta|<2 \sqrt6 a/3$ and $N\omega<\omega_{\textrm{max}}(=2\beta)$ where $N\geq2$ and $N\in\mathbb{N}$.

Let us now analyse the MI growth rate as a function of $\omega$ and $\beta$ in the focusing case, $\sigma=1$. The areas of nonzero $G$ are shown in Fig. \ref{f1}(b). The growth rate varies from $0$ to $2$. There are two specific regions in this plot. (i) The X-shaped region limited by the red solid hyperbolic curves $\omega^2=4\beta^2+8a^2$ and four dashed straight lines $\omega^2=4\beta^2$ limited by the condition $|{\omega}|>\sqrt2 a$. (ii) Two U-shaped regions limited by the two grey curves defined by $\omega^2=4a^2-a^4/\beta^2$.

In the region (i), the AB solutions satisfy the relation (\ref{eqpin}).
Thus, there is one AB solution in this area.
%Thus, these ABs are degenerate solutions.
On the other hand, in the region (ii) the eigenvalues are related by $\bm{\chi}_{+}=-\bm{\chi}_{-}-\omega$. Their imaginary parts are equal and have opposite sign, $\bm{\chi}_{+,i}=-\bm{\chi}_{-,i}$, but their real parts differ $\bm{\chi}_{+,r}\neq\bm{\chi}_{-,r}$. This means that the two solutions are different, $\psi^{(j)}(\bm{\chi}_{+})\neq\psi^{(j)}(\bm{\chi}_{-})$, despite   they have equal growth rates $G(\bm{\chi}_{+})=G(\bm{\chi}_{-})$. The two solutions, $\psi^{(j)}(\bm{\chi}_{+})$ and $\psi^{(j)}(\bm{\chi}_{-})$, do not satisfy the transformation (\ref{eqpin}). Thus, these two  AB solutions have the same initial growth rate but their wave profiles for any given values $a$, $\beta$, and $\omega$ are different.
Each of these ABs describes a distinctly individual growth-decay cycle of MI,
as shown in our previous work \cite{VAB2022},

% Figure 2
\begin{figure}[!htb]
\centering
\includegraphics[width=86mm]{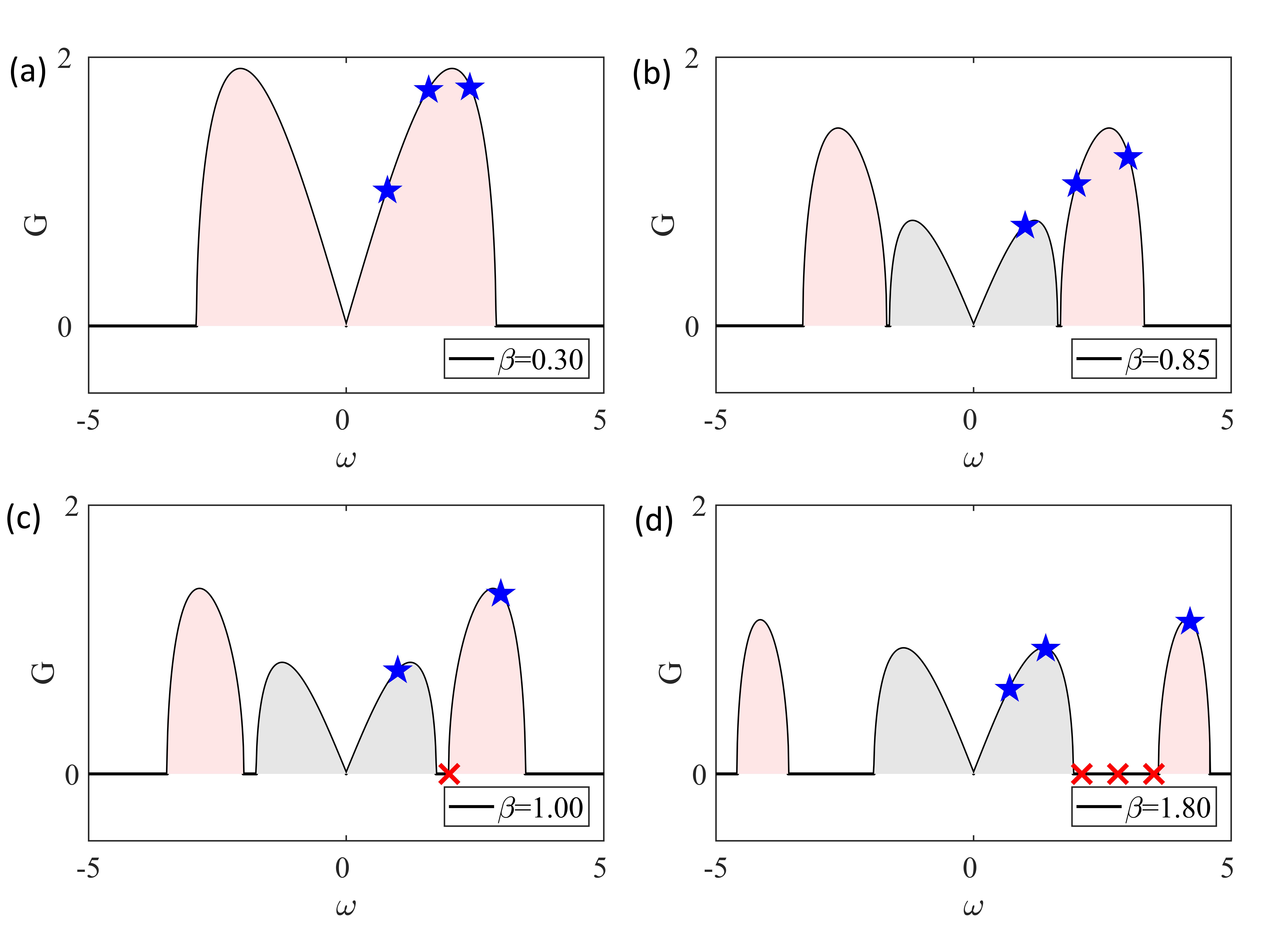}
\caption{Four examples of the MI growth rate spectra for fixed values of $\beta$ in the focusing case. Pink areas correspond to a single AB solution at each frequency. Grey areas contain two different AB solutions at each frequency. In all cases, $a=1$. \\
(a) $\beta=0.3$ ($\beta<|\beta_b|$). There is only one lobe at each side of the spectrum. The blue stars correspond to unstable modes at $\{\omega, 2\omega, 3\omega\}$ when $\omega=0.8$. \\
(b) $\beta=0.85$. There are two branches of instability at each side of the spectrum.
The blue star in the grey area corresponds to $\omega=1.0$ while the blue stars in the pink area correspond to $\{2\omega, 3\omega\}$. \\
(c) $\beta=1.0$. The two lobes of the MI growth rate at each side of the spectrum are separated with a small gap. The blue stars show the unstable mode at the first sideband $\omega=1$ in the grey area  and one unstable mode at the third harmonic $3\omega$ in the pink area. The perturbation at second harmonic $2\omega$ shown by the red cross does not grow.\\
(d) $\beta=1.8$. There are two unstable frequencies $\{\omega, 2\omega\}$ $(\omega=0.7)$ in the grey region and one with $\{6\omega\}$ in the pink region.}\label{f2}
\end{figure}

The four branch points shown by four large violet dots in Fig.\ref{f1}(b) are given by
\begin{eqnarray}
(\omega,\beta)=(\omega_b,\beta_b)=(\pm\sqrt{2}a,\pm\sqrt{2}a/2).\label{eq-bp}
\end{eqnarray}
In the horizontal stripe between the branch points, $\beta^2\leq\beta_b^2$. Here,
the MI growth rate has a single lobe at each side of the spectrum. Figure \ref{f2}(a) shows, as an example, the spectrum of the MI growth rate when $\beta=0.3$.
However, outside of the stripe, when $\beta^2>\beta_b^2$, the MI growth rate spectrum splits into two lobes at each side of the spectrum with the region of stability between them.
Three typical cases are shown in Figs. \ref{f2}(b)-\ref{f2}(d). Figure \ref{f2}(b) shows the MI growth rate spectrum when $\beta=0.85$ which is slightly higher than $\beta_b$. There is a small gap in the growth rate spectrum between the two lobes.
Figures \ref{f2}(c) and (d) show the MI growth rate spectra for larger values of $\beta$: $\beta=1.0$ and $\beta=1.8$ respectively.
The finite gap in the growth rate spectrum increases with $\beta$ as can be seen from these two examples.

In experiments, the first sideband of the induced MI spectrum can be chosen arbitrarily.
The choice of the first sideband influences the full scale evolution that starts with the MI. To give an example, the first sideband in Fig. \ref{f2}(a) is chosen to be $\omega=0.8$.
In this case, the two higher harmonics of the first sideband with frequencies $2\omega$, $3\omega$ are also within the MI band. They are shown in Fig. \ref{f2}(a) by the blue stars. Then, the higher-order MI will involve these three frequencies ($\omega$, $2\omega$, $3\omega$). All three of them initially will grow exponentially. The full scale evolution then will be described by the multi-AB solution with these frequencies. This dynamics is similar to the higher-order MI evolution in the scalar NLSE case \cite{Exp2011-PRL}. We omit this case in numerical simulations of the higher-order MI evolution.

The case shown in Fig. \ref{f2}(b) is more complicated. The MI growth rate spectrum now consists of the two lobes at each side of the symmetric spectrum. There are two different AB solutions at each frequency of the grey lobe of the spectrum. Both of them can be excited when the perturbation contains this frequency.
The pink lobe corresponds to a single AB at each frequency. Thus, only
 one AB can be excited at each of these frequencies.
 Let us suppose that the perturbation frequency is chosen to be $\omega=1$.
It is located in the grey lobe of the spectrum.  Then the two higher harmonics $2\omega$, and $3\omega$ are located on the pink lobe. These frequencies are shown by the blue stars in Fig. \ref{f2}(b). In this case, two ABs can be excited at the first sideband $\omega$ and two more ABs can be excited at frequencies $2\omega$, and $3\omega$. Thus, the full scale evolution of the induced MI with this frequency, $\omega=1$, will involve four ABs.

The gap between the spectral lobes increases with $\beta$. When this happens, one of the higher harmonics may fall into the gap and remain stable. Such case is shown in Fig. \ref{f2}(c). Here, $\beta=1.0$ and the first sideband is also chosen to be $\omega=1$. It is located slightly to the left of the maximum of the grey lobe of the spectrum. There are two different ABs that correspond to this frequency. The second harmonic, $2\omega$, appears in the gap between the two lobes. It is shown by the red cross in Fig. \ref{f2}(c). There is no growing AB at this frequency. On the contrary, the third harmonic $3\omega$ falls slightly to the right of the maximum of the pink lobe of the spectrum. It is
shown by the blue star in Fig. \ref{f2}(c). Thus, this frequency component is unstable and the AB with the corresponding frequency can be excited.
Thus, in the case of the induced MI with the frequency $\omega$, two ABs can be excited at the frequency $\omega$ and one AB at the frequency $3\omega$. There is no AB at the frequency $2\omega$. The full scale MI evolution will involve three ABs.

In the case shown in Fig. \ref{f2}(d), the value of $\beta$ is even higher ($\beta=1,8$). This leads to significantly wider gap between the two spectral lobes. Then, several higher harmonics of the first sideband may fall into this gap. For this to happen, the basic MI frequency is chosen to be $\omega=0.7$. This frequency and its second harmonic $2\omega$ are located in the grey lobe. They are shown by the blue stars on the grey lobe. The latter is close to the maximum of the growth rate. However, the three higher MI modes with the frequencies $3\omega$, $4\omega$ and $5\omega$ fall into the spectral gap. They are shown by the red crosses in Fig. \ref{f2}(d). There are no growing ABs at these frequencies. On the contrary, the sixth harmonic, $6\omega$, appears close to the maximum of the pink lobe. It is shown by the blue star.  It is unstable and the corresponding AB solution does exist. All together, four ABs can be excited within the grey lobe of the spectrum and
one AB can be excited in the pink area. Thus, the full scale evolution will involve five ABs.

When the modulation frequency is chosen as in the above examples, the induced MI process creates the higher harmonics of the first sideband. Those that are unstable will be amplified and generate the ABs. As discussed above, the number of excited ABs depends on the value of $\beta$. Each unstable frequency in the grey spectral lobes creates two ABs while each unstable frequency in the pink lobes creates one AB. This leads to the complex higher-order MI dynamics that can be described analytically using the higher-order AB solutions.
  For example, as four ABs can be excited in the case shown in Fig. \ref{f2}(b), this case can be described analytically with the fourth-order AB solution. Third-order AB solution is required for the case shown in Fig. \ref{f2}(c). The most complex case shown in Fig. \ref{f2}(d) requires fifth-order AB solution.

Our preliminary analysis shows that the induced higher-order MI may exhibit new complex dynamics which is absent in the scalar NLSE case \cite{Exp2011-PRL}.
Below, we provide numerical analysis of such dynamics confirming the above ideas.
It is based on direct simulations of Manakov equations with the MI related initial conditions.
We will also compare the numerical results with the exact multi-AB solutions of the same equations.

% Section III
\section{Higher-order vector MI dynamics}\label{Sec3}

\begin{figure*}[!htb]
\centering
\includegraphics[width=130mm]{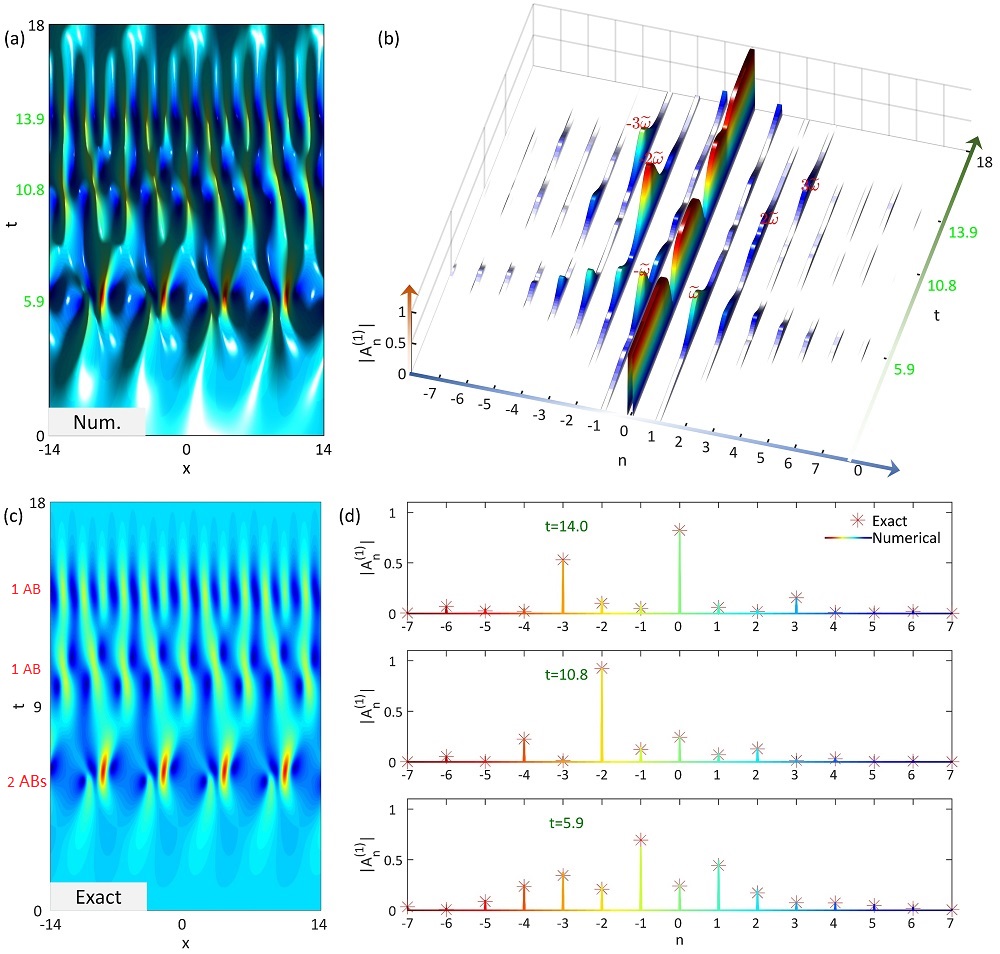}
\caption{Higher-order MI evolution with the growth rate spectrum shown in Fig. \ref{f2}(b).
Parameters $\tilde{\omega} = 1$, and $\beta = 0.85$.
(a) The results of numerical simulations started from initial conditions \eqref{eqin}. Only the $|\psi^{(1)}|$ wave component is shown. (b) The evolution of the discrete spectrum for the same simulations.
 (c) The exact 4th-order AB solution confirming numerical simulations. Parameters of the solution are: $\bm\chi_{\textrm{set}}=\{\bm\chi_{+}, \bm\chi_{-}, \bm\chi_{+}, \bm\chi_{+}\}$, $\omega_{\textrm{set}}=\{\tilde{\omega}, \tilde{\omega}, 2\tilde{\omega}, 3\tilde{\omega}\}$, $\Delta x_{\textrm{set}}=\{3.50, 0.90, -0.78, 0.78\}$ and $\Delta t_{\textrm{set}}=\{7.05, 5.80, 8.80, 11.95\}$.
 (d) Discrete spectra of the wave field obtained from the numerical simulations (vertical bars)  and from the exact solution (crosses) at three selected values of $t$ shown in green fonts in (a) and (b).
} \label{f3}
\end{figure*}

\begin{figure*}[htb]
\centering
\includegraphics[width=130mm]{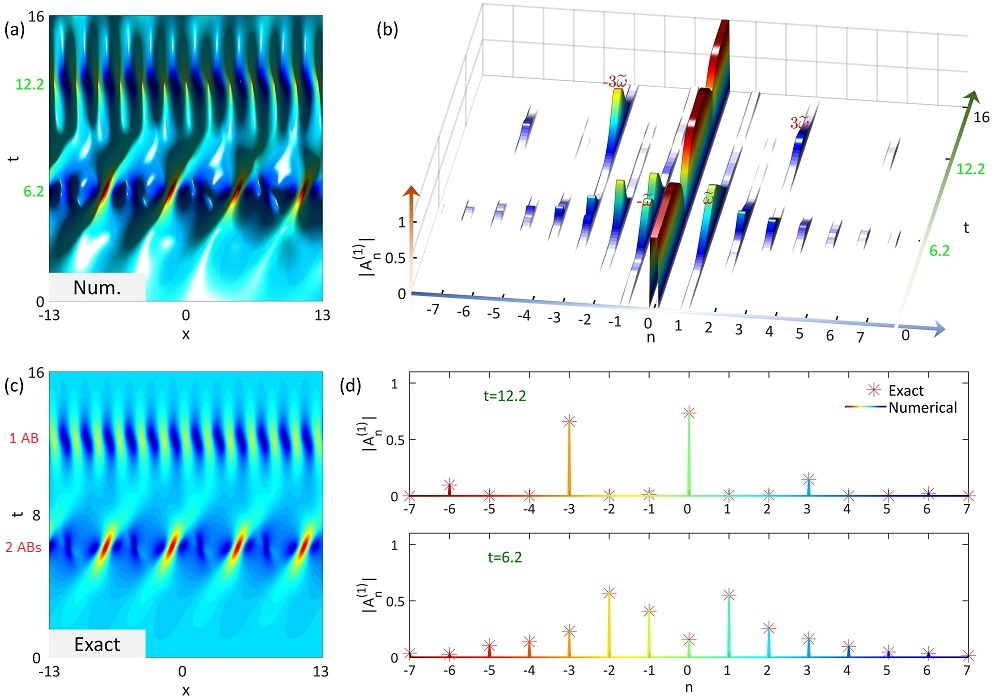}
\caption{Higher-order MI evolution with the growth rate spectrum shown in Fig. \ref{f2}(c).
Parameters $\tilde{\omega} = 1$, and $\beta = 1$. \\ (a) The results of numerical simulations started from initial conditions \eqref{eqin}. Only the $|\psi^{(1)}|$ wave component is shown. (b) The evolution of the discrete spectrum for the same simulations.
 (c) The exact 3th-order AB solution confirming numerical simulations. Parameters of the solution are: $\bm\chi_{\textrm{set}}=\{\bm\chi_{+}, \bm\chi_{-}, \bm\chi_{+}\}$ , $\omega_{\textrm{set}}=\{\tilde{\omega}, \tilde{\omega}, 3\tilde{\omega}\}$, $\Delta x_{\textrm{set}}=\{-3.29, -4.05, -0.62\}$ and $\Delta t_{\textrm{set}}=\{7.05, 5.90, 10.35\}$.
  (d) Discrete spectra of the wave field obtained from the numerical simulations (vertical bars)  and from the exact solution (crosses) at three selected values of $t$ shown in green fonts in (a) and (b).
} \label{f4}
\end{figure*}

\begin{figure*}[!htb]
\centering
\includegraphics[width=130mm]{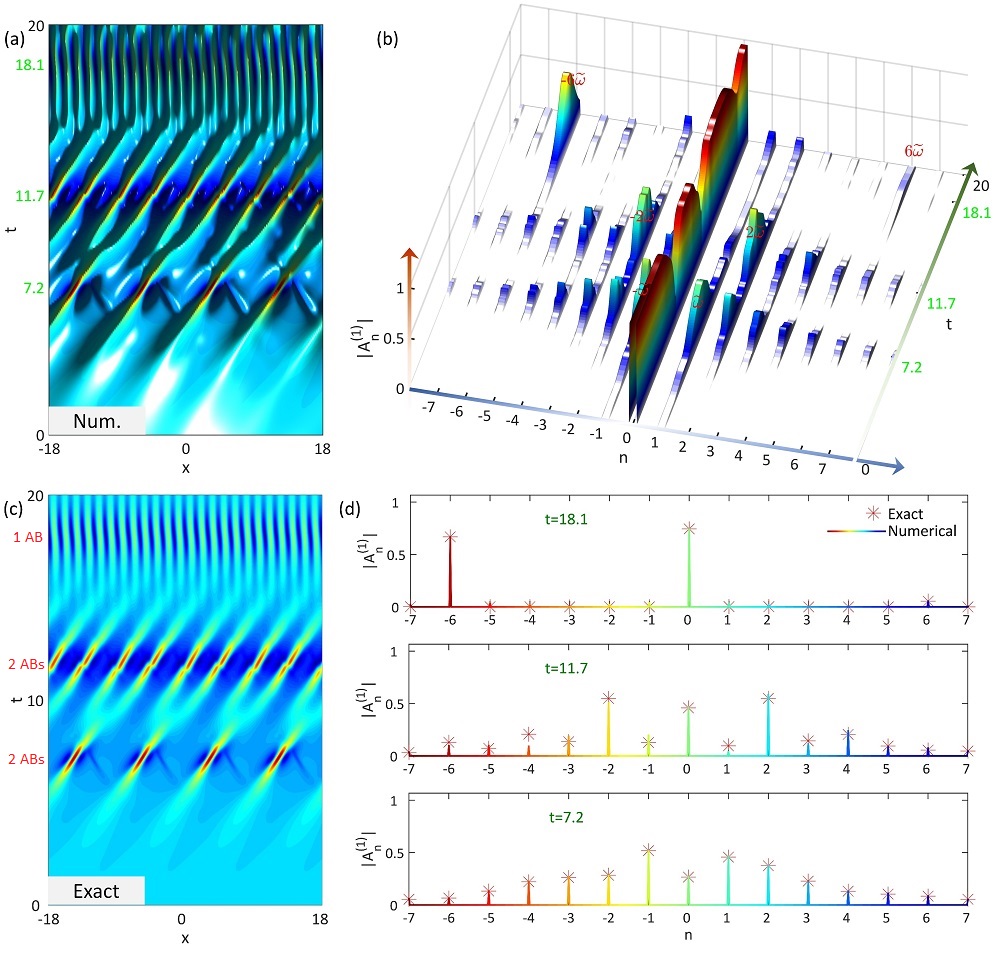}
\caption{Higher-order MI evolution with the growth rate spectrum shown in Fig. \ref{f2}(d).
Parameters $\tilde{\omega} = 0.7$, $\beta = 1.8$.
(a) The results of numerical simulations started from initial conditions \eqref{eqin}. Only the $|\psi^{(1)}|$ wave component is shown. (b) The evolution of the discrete spectrum for the same simulations.
(c) The exact 5th-order AB solution confirming numerical simulations. Parameters of the solution are: $\bm\chi_{\textrm{set}}=\{\bm\chi_{+}, \bm\chi_{-}, \bm\chi_{+}, \bm\chi_{-}, \bm\chi_{+}\}$ , $\omega_{\textrm{set}}=\{\tilde{\omega}, \tilde{\omega}, 2\tilde{\omega}, 2\tilde{\omega}, 6\tilde{\omega}\}$, $\Delta x_{\textrm{set}}=\{-1.88, -7.97, -0.42, -0.86, 0.34\}$ and $\Delta t_{\textrm{set}}=\{8.45, 8.20, 11.30, 10.90, 16.00\}$.
(d) Discrete spectra of the wave field obtained from the numerical simulations (vertical bars)  and from the exact solution (crosses) at three selected values of $t$ shown in green fonts in (a) and (b).
} \label{f5}
\end{figure*}

We simulated wave evolution by integrating numerically the set of Manakov equations (\ref{eq1}).
In simulations, we used the initial condition in the form of harmonically perturbed plane waves in each wave component:
\begin{equation}
\psi^{(j)}=(1+\varepsilon\cos\tilde{\omega} x)\psi_{0}^{(j)},\label{eqin}
\end{equation}
where $\varepsilon$ ($\ll1$) denotes a small amplitude of modulation with frequency $\tilde{\omega}$ ($\omega=\tilde{\omega}$), and $\psi_{0}^{(j)}$ is the plane wave background (\ref{eqb}). We accompanied these simulations with the analytic vector AB theory. The details of derivation of the higher-order AB solutions of a given order are presented in Appendix \ref{multi-AB}.
Parameters of the ABs in the theory have been chosen for best matching of the analytic results  with simulations. The results of the numerical simulations and the corresponding exact solutions are shown in Figs. \ref{f3}-\ref{f5}. The modulation frequencies and the growth rate spectra  chosen here are the same as in Figs. \ref{f2}(b)-\ref{f2}(d). We only show the MI dynamics in the $\psi^{(1)}$ component as the second one satisfies
the symmetry $\psi^{(2)}(\beta)=\psi^{(1)}(-\beta)$.

Figure \ref{f3} shows the higher-order MI dynamics when the spectrum of the MI growth rate and the modulation frequency are the same as in Fig. \ref{f2}(b). The results of numerical simulations for the evolution of the wave profile are shown in Fig. \ref{f3}(a). The corresponding evolution of the spectrum is shown in Fig. \ref{f3}(b).
The components of the discrete spectrum are numbered by the integer $n$.
The number $n=0$ corresponds to the pump mode while $|n|\geq1$ numbers the sidebands [$n=\pm1, \pm2, ... \pm7$]. The discrete spectral components at three selected values of $t$ are presented in Fig. \ref{f3}(d) as vertical bars. The selected values of $t$ are shown in green fonts in Figs. \ref{f3}(a) and \ref{f3}(b). The choice of these values of $t$ for presentation in Fig. \ref{f3}(d) is dictated by the points of maximal energy transfer from the pump to the sidebands.  These figures contain three growth-return cycles of the higher-order MI.
Figures \ref{f3}(b) and \ref{f3}(d) show that the l.h.s and r.h.s. sidebands are asymmetric with respect to the pump mode. This asymmetry indicates that there is transverse energy transfer in each wave component.

% Figure 3

As discussed above, the complex dynamics of the higher-order MI shown in Figs. \ref{f3}(a) and \ref{f3}(b) involves four fundamental ABs. Thus, the verification of these plots requires the fourth-order exact AB solution. This solution in matrix form is given in the Appendix.
The evolution of the wave profile $\psi^{(1)}$ according to this solution is shown in Fig. \ref{f3}(c). When parameters of the individual ABs are correctly chosen, the exact fourth-order AB solution reproduces well the numerical results.

The first growth-decay cycle is mainly defined by the two ABs $\psi^{(j)}(\bm{\chi}_{+}, \tilde{\omega})$ and $\psi^{(j)}(\bm{\chi}_{-}, \tilde{\omega})$, each with the transverse frequency $\omega=\tilde{\omega}$.
The two subsequent growth-decay cycles involve two ABs with the transverse frequencies
$k\tilde{\omega}$ where $k=2$ and $3$. These are $\psi^{(j)}(\bm{\chi}_{+}, 2\tilde{\omega})$ and $\psi^{(j)}(\bm{\chi}_{+}, 3\tilde{\omega})$.
Thus, the whole three-cycle dynamics of the higher-order MI contains four ABs expanding the spectrum from $\omega$, to $2\omega$, $3\omega$ and all higher harmonics of the modulation.

We also calculated the spectrum of the exact 4-th order AB solution using the method proposed in \cite{LA2021,VAB2021}. The discrete spectral components at the same three selected values of $t$ are shown by stars in Fig. \ref{f3}(d). As we can see, these stars are located on top of the corresponding bars obtained in numerical simulations. Thus, comparison of the spectra obtained in numerical simulations and from the exact solutions also shows good agreement between them.

% Figure 4

% Figure 5

Let us now turn our attention to the higher-order MI dynamics when some of the spectral components are located in stable regions of the growth rate spectrum as shown in Figs. \ref{f2}(c) and \ref{f2}(d). These spectral components do not excite the ABs.
In particular, Figure \ref{f4} shows the higher-order MI dynamics obtained for the growth rate spectrum in Fig. \ref{f2}(c).
Numerical results for the wave evolution excited by the initial conditions (\ref{eqin}) with $\tilde{\omega}=1$ and $\beta=1$ are displayed in Figs. \ref{f4}(a).
The corresponding spectrum is shown in Fig. \ref{f4}(b).
The higher-order MI in this example exhibits only two growth-decay cycles.
The initial modulation in Fig. \ref{f4}(a) develops into the ABs with the transverse period $2\pi/\tilde{\omega}$ and the maximum modulation at $t=6.2$.
This structure further evolves into a breather with the period $2\pi/(3\tilde{\omega})$.
Each maximum of the previous breather splits into the three smaller maxima rather than two.
Correspondingly, the spectrum evolution shown in Fig. \ref{f4}(b) reveals the enhancement of the third-order sidebands ($\pm3\tilde{\omega}$) at the second expansion-contraction cycle.
Instead, the second-order sidebands ($\pm2\tilde{\omega}$) are completely suppressed.

The first cycle consists of two different ABs with the same frequency $\tilde{\omega}$, namely $\psi^{(j)}(\bm{\chi}_{+}, \tilde{\omega})$ and $\psi^{(j)}(\bm{\chi}_{-}, \tilde{\omega})$. The second cycle can be described by the AB $\psi^{(j)}(\bm{\chi}_{+}, 3\tilde{\omega})$. This corresponds to the breather evolution with abnormal frequency jumping ($\tilde{\omega}\rightarrow3\tilde{\omega}$).
Figure \ref{f4}(c) shows the exact third-order AB solution formed by nonlinear superposition between $\psi^{(j)}(\bm{\chi}_{+}, \tilde{\omega})$, $\psi^{(j)}(\bm{\chi}_{-}, \tilde{\omega})$ and $\psi^{(j)}(\bm{\chi}_{+}, 3\tilde{\omega})$. The exact results in Fig. \ref{f4}(c) are in good agreement with the numerical simulations in Fig. \ref{f4}(a). The discrete spectra obtained from the numerical simulations (vertical bars) and the exact results (stars) at selected values of $t$ are shown in Fig. \ref{f4}(d). The selected values of $t$ here correspond to the maximal energy transfer from the carrier wave to the sidebands. Comparison of these spectra also shows good agreement between the numerical simulations and analytic results.

Figure \ref{f5} shows the vector higher-order MI corresponding to the MI growth rate shown in Fig. \ref{f2}(d).
Figures \ref{f5}(a) and \ref{f5}(b) show the numerical results of the evolution in time and frequency domains, respectively.
Three growth-return cycles of the amplitude distribution can be seen from Fig. \ref{f5}(a).
Specifically, the initial modulation grows exponentially into breathers with period $2\pi/\tilde{\omega}$.
The breathers then split into subwaves with period $\pi/\tilde{\omega}$.
After that, these subwaves split into small-amplitude breathers with smaller period $\pi/(3\tilde{\omega})$.
As the breathers of the first two cycles exist in the
%non-degenerate
region (ii), each cycle consists of two different ABs. Namely, the first cycle corresponds to the
%non-degenerate
ABs $\{\psi^{(j)}(\bm{\chi}_{+}, \tilde{\omega}), \psi^{(j)}(\bm{\chi}_{-}, \tilde{\omega})\}$; the second cycle
consists of the
%non-degenerate
ABs $\{\psi^{(j)}(\bm{\chi}_{+}, 2\tilde{\omega}), \psi^{(j)}(\bm{\chi}_{-}, 2\tilde{\omega})\}$.
This corresponds to the breather splitting with normal frequency jumping ($\tilde{\omega}\rightarrow2\tilde{\omega}$) in the
%non-degenerate
region (ii).
On the other hand, the breather of the third cycle corresponds to the
%degenerate
region (i), which is given by either $\psi^{(j)}(\bm{\chi}_{+}, 6\tilde{\omega})$ or $\psi^{(j)}(\bm{\chi}_{-}, 6\tilde{\omega})$. This corresponds to the breather splitting with abnormal frequency jumping ($2\tilde{\omega}\rightarrow6\tilde{\omega}$). This can be confirmed by the spectrum evolution shown in Fig. \ref{f5}(b). As can be seen, only the $n=-6$ sideband is well enhanced at the third expansion-contraction cycle.

% Figure 6
\begin{figure}[htb]
\centering
\includegraphics[width=86mm]{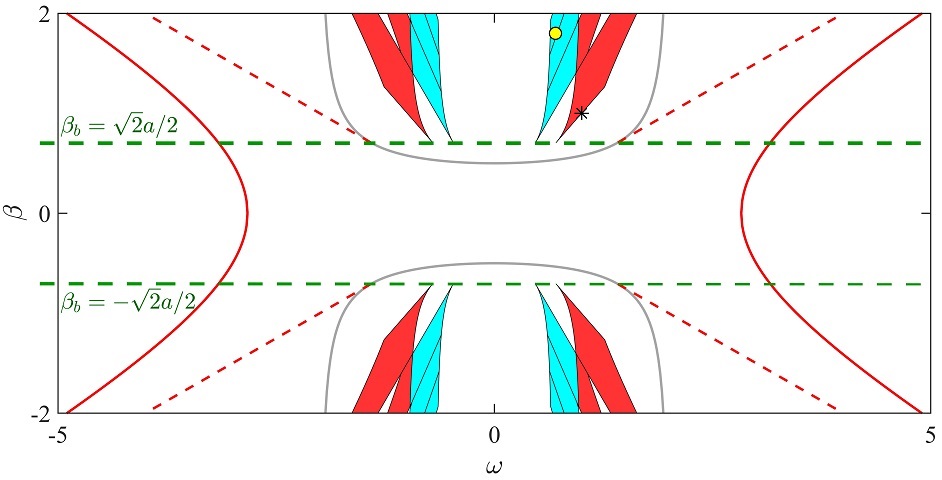}
\caption{The ($\omega,\beta$) plane of initial conditions.  The higher-order MI with two types of frequency jumping can be excited in the cyan and red areas.
The black star corresponds to the initial parameters that we used to generate higher-order MI dynamics shown in Fig. \ref{f4} while the yellow dot shows the parameters used for generation of results shown in Fig. \ref{f5}.}\label{f6}
\end{figure}

Just as in the two other cases discussed above, the higher-order MI dynamics involving three growth-return cycles shown in Fig. \ref{f5}(a) can be well described by the exact higher-order AB solution. Figure \ref{f5}(c) shows the amplitude distribution of the exact solution formed by the nonlinear superposition of five ABs. The corresponding spectra at the selected time are shown in Fig. \ref{f5}(d).
Again, there is great agreement between the exact solution and the numerical simulations.

Results of higher-order MI dynamics shown in Figs. \ref{f4} and \ref{f5} are only particular examples involving multi-ABs with abnormal frequency jumping.
In fact, we have found that such MI dynamics can be excited from a wide range of the parameters of initial conditions \eqref{eqin}. In order to prove this point, in Fig. \ref{f6}, we present the existence diagram of such excitations on the ($\omega,\beta$) plane obtained numerically.
Such higher-order MI dynamics can be excited when the parameters $\omega$, and $\beta$ are in the cyan and red areas. They belong to the low-frequency MI subregion with $\beta>|\beta_b|$.
Higher-order MI in the red area involves the energy transfer between the spectral components that are not the nearest neighbours, i.e. ($\tilde{\omega}\rightarrow k\tilde{\omega}$), where $k\geq3$. A particular case with $k=3$ is shown in Fig. \ref{f4}.
The parameters of the initial conditions that correspond to this case are represented by the black star in Fig. \ref{f6}.

On the other hand, higher-order MI in the cyan area involves, as the first step, energy transfer between the closest components ($\tilde{\omega}\rightarrow2\tilde{\omega}$). The second step is the energy transfer across the spectral components:  ($2\tilde{\omega}\rightarrow k\tilde{\omega}$) with  $k\geq4$.
The longer spectral jump when $k=6$ is shown in Fig. \ref{f5}. The parameters of the initial conditions that correspond to this case are represented by the yellow dot in Fig. \ref{f6}.
The sharp edges of the red and cyan regions in Fig. \ref{f6} are located on the green dashed line  $\beta=\beta_b=\pm\sqrt{2}a/2$. The diagram in Fig. \ref{f6} can be useful in experiments for observations of higher-order MI in optics and hydrodynamics.

% Section IV
\section{Conclusions}\label{}

In conclusion, we have studied higher-order MI dynamics that start with a single frequency modulation using direct numerical simulations of the Manakov equations. Due to the existence of additional  modulation instability bands in the case of the Manakov equations, the nonlinear stage of the modulation instability is qualitatively different from the MI of the single NLSE. We presented detailed analysis of the processes in such dynamics and confirmed the analysis using the exact multi-AB solutions of Manakov equations. We have shown that the energy of the spectral components of the vector higher-order MI may jump over the frequency regions that belong to the stable gaps of the growth rate spectrum.

\section*{ACKNOWLEDGEMENTS}
The work of Liu is supported by the NSFC (Grants No. 12175178, and No. 12047502),
the Natural Science basic Research Program of Shaanxi Province (Grant No. 2022KJXX-71), and Shaanxi Fundamental
Science Research Project for Mathematics and Physics (Grant No. 22JSY016).
The work of Akhmediev is supported by the Qatar National Research Fund (grant NPRP13S-0121-200126).
\begin{appendix}

\section{multi-AB solutions}\label{multi-AB}
The general determinant form of the $M$-th order AB solutions via the  B\"{a}cklunk transformation is:
\begin{eqnarray}
\psi^{(j)}[M]=\psi_{0}^{(j)}{\det(\mathcal{G}^{(j)})}/{\det(\mathcal{G})},\label{vg1}
\end{eqnarray}
where
\begin{eqnarray}
\mathcal{G}^{(j)}&=&\begin{pmatrix} g^{(j)}_{1,1} & g^{(j)}_{1,2}& ... & g^{(j)}_{1,M}\\
g^{(j)}_{2,1} & g^{(j)}_{2,2}& ... & g^{(j)}_{2,M}\\
\vdots&\vdots& &\vdots&\\
g^{(j)}_{M,1} & g^{(j)}_{M,2}& ... & g^{(j)}_{M,M} \end{pmatrix},\\
\mathcal{G}&=&\begin{pmatrix} g_{1,1} & g_{1,2}& ... & g_{1,M}\\
g_{2,1} & g_{2,2}& ... & g_{2,M}\\
\vdots&\vdots& &\vdots&\\
g_{M,1} & g_{M,2}& ... & g_{M,M} \end{pmatrix}.
\end{eqnarray}
Here, $g^{(j)}_{m1,m2}$ and $g_{m1,m2}$ are the matrix elements of $\mathcal{G}^{(j)}$ and $\mathcal{G}$ in the $m1$-th row, $m2$-th column, respectively.
They are given by:
\begin{eqnarray}\label{ys}
\begin{split}
g_{m1,m2}&=\frac{\varphi_{m1}\varphi^*_{m2}}{\bm\chi^*_{m2}-\bm\chi_{m1}}+
\frac{\tilde{\varphi}_{m1}\tilde{\varphi}^*_{m2}}{\tilde{\bm\chi}^*_{m2}-\tilde{\bm\chi}_{m1}}
+\frac{\varphi_{m1}\tilde{\varphi}^*_{m2}}{\tilde{\bm\chi}^*_{m2}-\bm\chi_{m1}}\\
&+\frac{\tilde{\varphi}_{m1}+\varphi^*_{m2}}{\bm\chi^*_{m2}-\tilde{\bm\chi}_{m1}},\nonumber\\
g^{(j)}_{m1,m2}&=\frac{\bm\chi^*_{m2}+\beta_j}{\bm\chi_{m1}+\beta_j}
\frac{\varphi_{m1}\varphi^*_{m2}}{\bm\chi^*_{m2}-\bm\chi_{m1}}+
\frac{\tilde{\bm\chi}^*_{m2}+\beta_j}{\tilde{\bm\chi}_{m1}+\beta_j}
\frac{\tilde{\varphi}_{m1}\tilde{\varphi}^*_{m2}}{\tilde{\bm\chi}^*_{m2}-\tilde{\bm\chi}_{m1}}\\
&+\frac{\tilde{\bm\chi}^*_{m2}+\beta_j}{\bm\chi_{m1}+\beta_j}
\frac{\varphi_{m1}\tilde{\varphi}^*_{m2}}{\tilde{\bm\chi}^*_{m2}-\bm\chi_{m1}}+
\frac{\bm\chi^*_{m2}+\beta_j}{\tilde{\bm\chi}_{m1}+\beta_j}
\frac{\tilde{\varphi}_{m1}\varphi^*_{m2}}{\bm\chi^*_{m2}-\tilde{\bm\chi}_{m1}},
\end{split}
\end{eqnarray}
where $\ast$ denotes the complex conjugate, $\bm\chi_{m}$ is the eigenvalue and $\tilde{\bm\chi}_m=\bm\chi_m+\omega_m$ $(m=1, 2, 3, ... M)$. Note that \begin{eqnarray}
\bm\chi_{m1}=\bm\chi_{m}|_{m=m1},~\tilde{\bm\chi}_{m1}=\tilde{\bm\chi}_{m}|_{m=m1},\nonumber\\
\bm\chi_{m2}=\bm\chi_{m}|_{m=m2},~\tilde{\bm\chi}_{m2}=\tilde{\bm\chi}_{m}|_{m=m2}.\nonumber
\end{eqnarray}

Similarly,
\begin{eqnarray}
\varphi(\bm\chi_{m1})=\varphi(\bm\chi_{m})|_{m=m1},~\varphi(\tilde{\bm\chi}_{m1})=\varphi(\bm\chi_{m})|_{m=m1},\nonumber\\
\varphi(\bm\chi_{m2})=\varphi(\bm\chi_{m})|_{m=m1},~
\varphi(\tilde{\bm\chi}_{m2})=\varphi(\bm\chi_{m})|_{m=m2}.\nonumber
\end{eqnarray}
where the functions $\varphi(\bm\chi_m)$, and $\varphi(\tilde{\bm\chi}_m)$  are given by
\begin{eqnarray}
\varphi(\bm\chi_m)=\exp\{i\bm\chi_m[(x-x_{m})+\frac{1}{2}\bm\chi_m(t-t_{m})]\},\label{vg1}\\
\varphi(\tilde{\bm\chi}_m)=\exp\{i\tilde{\bm\chi}_m[(x-x_{m})+\frac{1}{2}\tilde{\bm\chi}_m(t-t_{m})]\}.\label{vg1}
\end{eqnarray}
The real parameters $x_m$, and $t_m$ are the shifts in $x$ and $t$ of individual breathers, respectively.
For $M=1$, we obtain the fundamental vector AB solution of the Manakov equations. It is given, in simplified form, by Eq. (\ref{eqb}).

The $M$th-order solution corresponds to the nonlinear superposition of $M$ fundamental ABs, each associated with the parameters $(\bm\chi_m, \omega_m, x_{m},t_{m})$, where $m=1,...,M$.
The space-time structure of a single AB in the superposition is directly determined by the parameters $\bm\chi_{\textrm{set}}=\{\bm\chi_{1}, ..., \bm\chi_{M}\}$ and the frequencies $\omega_{\textrm{set}}=\{\omega_{1}, ..., \omega_{M}\}$.
The interaction between them (the spatiotemporal patterns of such multi-ABs) depends on the relative separations in both $x$ and $t$, i.e., $\Delta x_{\textrm{set}}=\{x_{1}, ..., x_{M}\}$ and $\Delta t_{\textrm{set}}=\{t_{1}, ..., t_{M}\}$.

\end{appendix}

\end{CJK}

\end{document}